\documentclass[letterpaper, 10 pt, conference]{ieeeconf}  

\IEEEoverridecommandlockouts                              

\usepackage{graphics} 
\usepackage{epsfig} 
\usepackage{mathptmx} 
\usepackage{times} 
\usepackage{amsmath} 
\usepackage{amssymb}  
\usepackage{url}

\title{\LARGE \bf
Software Compensation of Undesirable Racking Motion of 
H-frame 3D Printers using Filtered B-Splines
}

\author{Nosakhare Edoimioya, Keval S. Ramani, Chinedum E. Okwudire
\thanks{Department of Mechanical Engineering,
        University of Michigan-Ann Arbor, 2350 Hayward Street, Ann Arbor, MI 48104, USA.
        {\tt\small \{nosed,ksramani,okwudire\}@umich.edu}}%
}

\begin{document}

\maketitle
\thispagestyle{empty}
\pagestyle{empty}

\begin{abstract}

The H-frame (also known as H-Bot) architecture is a simple and elegant two-axis parallel positioning system used to construct the XY stage of 3D printers. It holds potential for high speed and excellent dynamic performance due to the use of frame-mounted motors that reduce the moving mass of the printer while allowing for the use of (heavy) higher torque motors. However, the H-frame’s dynamic accuracy is limited during high-acceleration and high-speed motion due to racking – i.e., parasitic torsional motions of the printer’s gantry due to a force couple. Mechanical solutions to the racking problem are either costly or detract from the simplicity of the H-frame. 
In this paper, we introduce a feedforward software compensation algorithm, based on the filtered B-splines (FBS) method, that rectifies errors due to racking. The FBS approach expresses the motion command to the machine as a linear combination of B-splines. The B-splines are filtered through an identified model of the machine dynamics and the control points of the B-spline based motion command are optimized such that the tracking error is minimized. To compensate racking using the FBS algorithm, an accurate frequency response function of the racking motion is obtained and coupled to the H-frame’s $x$-  and $y$-axis dynamics with a kinematic model. The result is a coupled linear parameter varying model of the H-frame that is utilized in the FBS framework to compensate racking. An approximation of the proposed racking compensation algorithm, that decouples the $x$- and $y$-axis compensation, is developed to significantly improve its computational efficiency with almost no loss of compensation accuracy. Experiments on an H-frame 3D printer demonstrate a 43\% improvement in the shape accuracy of a printed part using the proposed algorithm compared to the standard FBS approach without racking compensation. The proposed racking compensation algorithm can be used in-conjunction with mechanical solutions, or as a stand-alone solution, to improve the performance of H-frame architectures.

\end{abstract}

\section{Introduction}

Fused deposition modeling (FDM) 3D printers, which represent approximately 75\% of the 3D printing market [1], manufacture parts by extruding material from a heated nozzle onto a bed, with the help of motion systems that move the nozzle and bed. The standard choice for motion systems of FDM 3D printers is the so-called serial stack architecture, which generates motion using independent (i.e., decoupled) actuators for each axis, and typically requires one of the axes and its associated motor(s) to be “stacked” on another axis. This stacking leads to high inertial loads and motion friction, which limit the dynamic accuracy and the available torque during high speed motions. Although the serial stack architecture is still a popular choice, less conventional motion systems such as the H-frame [2], Delta [3], and Core XY [4] architectures are designed using stationary motors to reduce the moving mass of the end-effector, allowing for increased dynamic accuracy and motion speed. 

The H-frame architecture, developed by Stratasys, Inc. in 2011 [2], found early success due to the simplicity of its parallel axis design. As seen in Fig. 1(a), it consists of two motors mounted to the frame of the 3D printer, which are connected to the end-effector through a single timing belt. Translational motion in the $x$- and $y$-axis of the end-effector are generated via the rotational motion of the frame-mounted motors, which is transmitted by a timing belt and pulley configuration [2,5,6]. Since the motors are stationary, high-power (and, typically, heavy) motors can be utilized to achieve high-speed, high-precision motion without increasing the moving mass of the printer. For this reason, the H-frame architecture has been used in several 3D printers, such as the Stratasys Mojo [7], MakerBot Replicator Z18 [8], Creality Ender 4 [9], and MIT’s Fast FFF [10] 3D printers. Using the H-frame as one of several improvements to conventional FDM 3D printers, Go et al. [10] demonstrated 5 to 10 times improvements in build rate using the Fast FFF printer compared to several printers of the same class. However, parts printed with H-frame 3D printers suffer from quality defects caused by parasitic error motions due to “racking” [11-13]: when the motors are commanded to rotate in the same direction, corresponding to x-axis motion of the end-effector, a force couple (pure moment) is imposed on the gantry (Fig. 1(b)) which, depending on the speed, may create large enough errors to distort the part shape. The magnitude of these errors also depends on the end-effector’s location along the $x$-axis, indicating a parameter varying system [13].

\begin{figure*}[]
	\centering		\includegraphics[width=.80\textwidth]{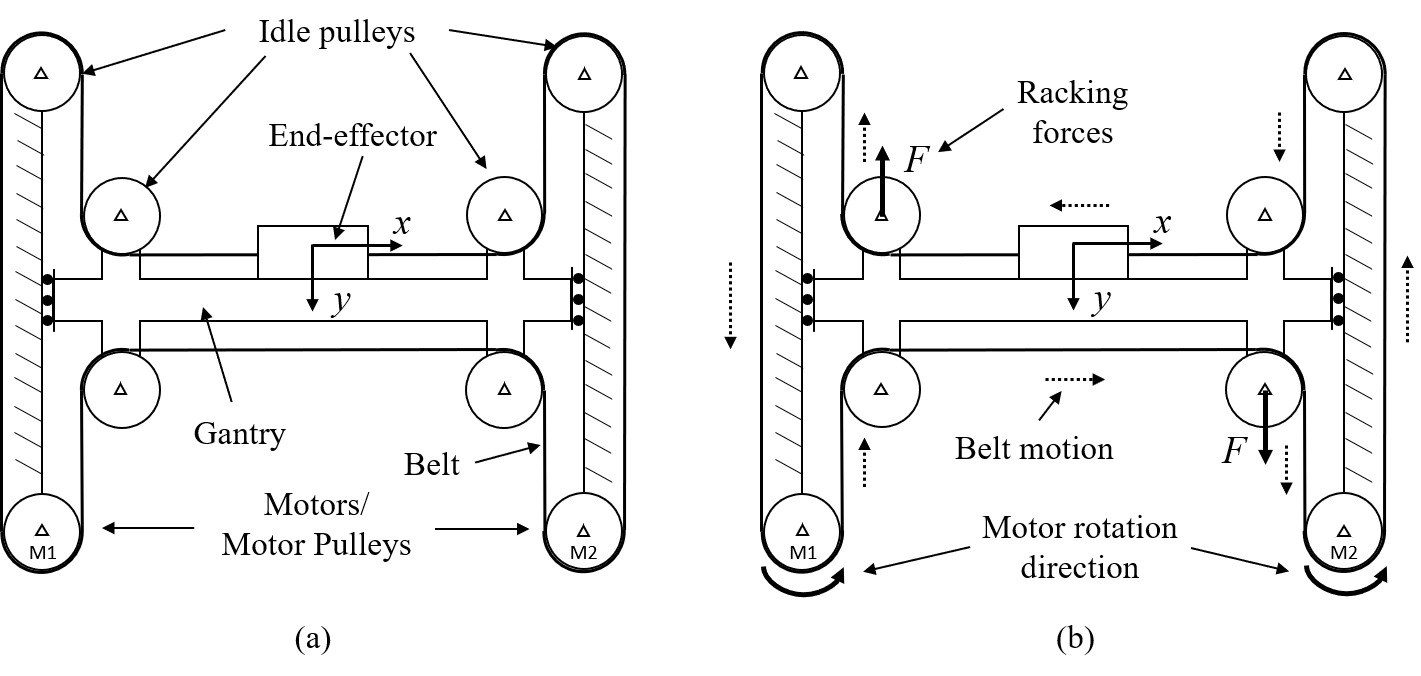}
    \caption{(a) The H-frame architecture with stationary motors labeled M1 and M2 and the timing belt and pulley configuration that transmits rotational motion to XY translational motion of the gantry and end-effector; (b) Unbalanced forces, $F$, that cause racking—parasitic torsional motions—of the gantry.}
    \label{fig:fig01}
\end{figure*}

Racking errors can be mitigated with mechanical solutions such as a rigid linear guideway design, which are typically expensive. A lower-cost option is to design a modified configuration, such as the two-belt Core XY architecture [4], which ensures the forces on the gantry do not create a force couple. However, the Core XY and similar designs can be significantly more complex than the H-frame’s design, and difficult to manufacture. Additionally, other sources of error may surface, such as the errors created on the Core XY when the two belts are not equally tensioned [13]. 

Conversely, software compensation (i.e., feedback and/or feedforward control algorithms) can be used to reduce or eliminate racking errors, often without need to modify the mechanical architecture of the printer. Although useful in a variety of applications, feedback (FB) control, which depends on sensing to correct errors, is impractical for racking compensation for the following reasons [6,14]: (1) It is not applicable to a wide range of 3D printers that are stepper motor-controlled in the open-loop (i.e., they have no position sensors); (2) the position sensors available on some 3D printers are motor-mounted, hence cannot sense racking at the end-effector; and (3) adding sensors that can sense racking at the end-effector will lead to non-collocated control systems that are prone to instability [15]. 

The challenges of FB control can be mitigated via feedforward (FF) control, which compensates errors using a model of the controlled system as opposed to sensing. Feedforward control has been shown to improve motion accuracy in a host of manufacturing applications, including 3D printing [16-24]. A popular FF control approach for reducing motion errors is called smooth command generation [21,25] where motion commands are generated to have little to no high-frequency content by using, for example, low-pass filters [26] and jerk-limited [25,27] trajectories. However, the attenuated high-frequency content of smooth command generation implies loss of motion speed, which adversely affects productivity [18]. Additionally, smooth command generation methods are sub-optimal because motion commands are not generated with knowledge of the machine’s dynamics [28]. Hence, conservative acceleration and jerk limits are often adopted in practice since there is no clear understanding of how to select the limits to achieve a desired performance metric [29]. Input shaping [30-32], another popular FF control method, eliminates vibration errors through destructive interference by commanding a series of impulses that are equal in magnitude but opposite in phase to the vibration errors of the system. A major limitation of input shaping is that it introduces time delays between the desired and actual motions, leading to large tracking/contouring errors and reduction of productivity [33].  Therefore, while it works well for point-to-point motions, it exhibits poor performance for the tracking/contouring motions prevalent in most AM applications. Another class of FF control methods is known as model-inversion based FF control. Methods in this class compensate motion errors by using the inverse of the motion system’s dynamics to pre-filter motion commands. Model inversion-based FF control methods do not introduce time delays and can theoretically lead to perfect compensation of motion errors [34]. There are several types of model inversion-based FF controllers available, as discussed extensively in [34-36]. Of the available methods, the filtered basis functions (FBF) approach has been shown to be very versatile, compared to others, with regards to its applicability to any linear system dynamics [16,17,37-40]. The FBF approach expresses motion commands as a linear combination of basis functions, forward filters the basis functions using the plant dynamics, and calculates the coefficients of the basis functions such that motion errors are minimized. A version of FBF commonly used for controlling manufacturing machines is the filtered B-splines (FBS) method [16-19,37] where B-splines are selected as the basis functions, because they are amenable to the lengthy motion trajectories common in manufacturing. The FBS method has been used to reduce the printing time of serial stack FDM 3D printers by up to 54\% without sacrificing print quality, by compensating vibration-induced errors [16,18,19]. However, the standard implementation of the FBS method, used in the aforementioned studies, is not directly applicable to the compensation of racking in H-frame 3D printers because it assumes that the printer dynamics is linear time invariant (LTI) and decoupled, whereas, racking dynamics is linear parameter varying (LPV) and coupled.

\begin{figure*}[ht]
	\centering		
	\includegraphics[width=.80\textwidth]{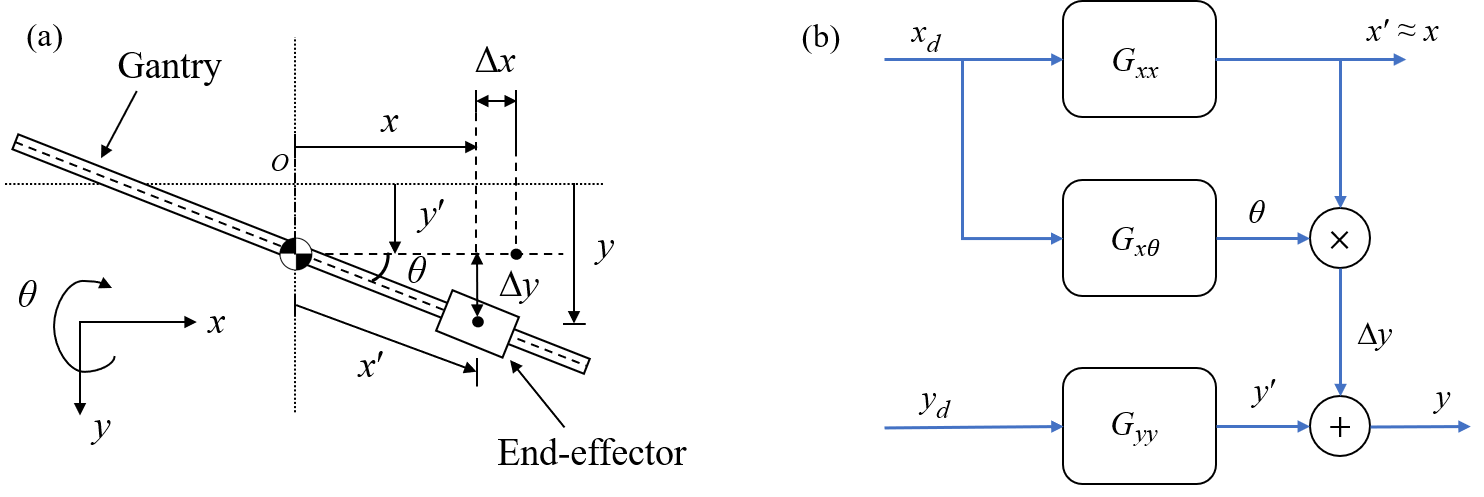}
    \caption{(a) Kinematic model of parasitic errors created by racking motions of H-frame 3D printers; (b) Dynamic model of H-frame 3D printers including effect of parasitic $y$ motions ($\Delta y$) created by racking.}
    \label{fig:fig02}
\end{figure*}

Therefore, to enable compensation of racking errors in H-frame 3D printers, this paper makes the following original contributions:
\begin{enumerate}
    \item It proposes an extension of the standard FBS controller designed to compensate the coupled LPV racking dynamics of H-frame 3D printers.
    \item It develops a simplification (i.e., decoupled version) of the designed coupled LPV FBS controller that significantly reduces its computational cost with little to no sacrifice to its racking compensation accuracy.
    \item It demonstrates the effectiveness and practicality of the developed algorithm in compensating the racking errors through simulations and experiments on an H-frame 3D printer.
\end{enumerate}
The outline of the paper is as follows: Section II introduces and validates a model of H-frame racking motion on an H-frame 3D printer. Section III gives an overview of the standard FBS method and then proposes a coupled LPV FBS method for compensating racking errors. Section III also demonstrates the increased computational cost of the proposed coupled LPV FBS controller, relative to the standard FBS controller, and proposes a simplification that reduces its computational cost—with minimal sacrifice to its performance—hence facilitating its practicality. Section IV presents simulations and experiments on the H-frame 3D printer modeled in Section 2, followed by conclusions in Section V.

\section{Model of Parasitic Racking Motion of H-frame}

The racking motion of the gantry on the H-frame is caused by a force couple (pure moment) which creates an angular displacement in rotational axis, $\theta$, on the gantry. Let $\{x, y\}$ be the end-effector’s location. It can be decomposed into two portions: $\{x',y'\}$, the position of the end-effector without the errors introduced by racking; and $\{\Delta x,\Delta y\}$ due to the error created by racking angle $\theta$  as shown in Fig. 2(a). Using the small angle approximation (i.e., $\cos\theta\approx 1$, $\sin\theta\approx\theta$), $\Delta x\approx0$ and we can write the $x$- and $y$-axis locations as:
\begin{align}
    x & = x'\cos\theta\approx x' \\
    y & = y' + \Delta y = y' + x'\sin\theta\approx y' + x\theta
\end{align}
This kinematic model can be used to create a coupled dynamic model including racking, as shown in the block diagram in Fig. 2(b), where $\{x_d,y_d\}$ represent the desired position of the end-effector. The transfer functions from $x_d$  to $x'$, and $y_d$ to $y'$ are represented by $G_{xx}$ and $G_{yy}$, respectively, whereas, $G_{x\theta}$ represents the racking contribution (i.e., the transfer function from $x_d$ to $\theta$).

In addition to the small angle approximation, two other assumptions are implied in the model of Fig. 2(b). The first is that the H-frame dynamics can be approximated as linear. Therefore, $x'$, $y'$, and $\theta$ can be derived from transfer functions $G_{xx}$, $G_{yy}$ and $G _{x\theta}$, respectively. This assumption has been found to be reasonable in prior work [5,6,14,16-19]. The second assumption is that the transfer function $G_{x\theta}$  does not vary as a function of end-effector position. This implies that the inertia of the gantry is not significantly affected by the position of the end-effector as will be validated later in this section.

\noindent \underline{\textit{Remark 2.1}}: The model shown in Fig. 2(b) is nonlinear because $\Delta y$  is generated from the product of two outputs. It can be approximated as linear parameter varying (LPV) by assuming that $x \approx x_d$ when determining $\Delta y$. This assumption is reasonable because the tracking errors, $x_d-x$, caused by $G_{xx}$ are typically much smaller than the magnitude of $x$. Therefore, they have insignificant contributions to $\Delta y$. Accordingly, $\Delta y = x_d\theta$ is assumed in the rest of this paper, for the sake of simplicity, resulting in a coupled LPV model for H-frame 3D printers.

The 3D printer shown in Fig. 3(a) is used to validate the H-frame model of Fig. 2. It is fabricated by adapting a Creality Ender 5 3D printer into its H-frame predecessor, the Ender 4. (The Ender 4 was discontinued by Creality, hence was unavailable for purchase.) The designed H-frame configuration is actuated by two NEMA 17 stepper motors via a 2-mm pitch, 6-mm wide, rubber timing belt used for motion transmission (more details in [41]). The motors are controlled by Pololu DRV8825 high-current stepper motor drivers configured to give a step resolution of 2$\mu$ rad per step which is transmitted through a pulley with radius $r = 5.15$ mm to give a $x$- and $y$-axis resolution of 20.6 $\mu$m per step. The motion range of the $x$ and $y$ axes are 280 and 295 mm, respectively. Real-time control of the $x$, $y$ and $z$ axes and extrusion motors is performed using dSPACE MicroLabBox (RTI 1202) with stepping frequency of 40 kHz and sampling frequency of 1 kHz. Commands to the printer were generated in MATLAB and sent to the MicroLabBox through a MATLAB Simulink interface. 

\begin{figure}[]
	\centering
    \includegraphics[width=.42\textwidth]{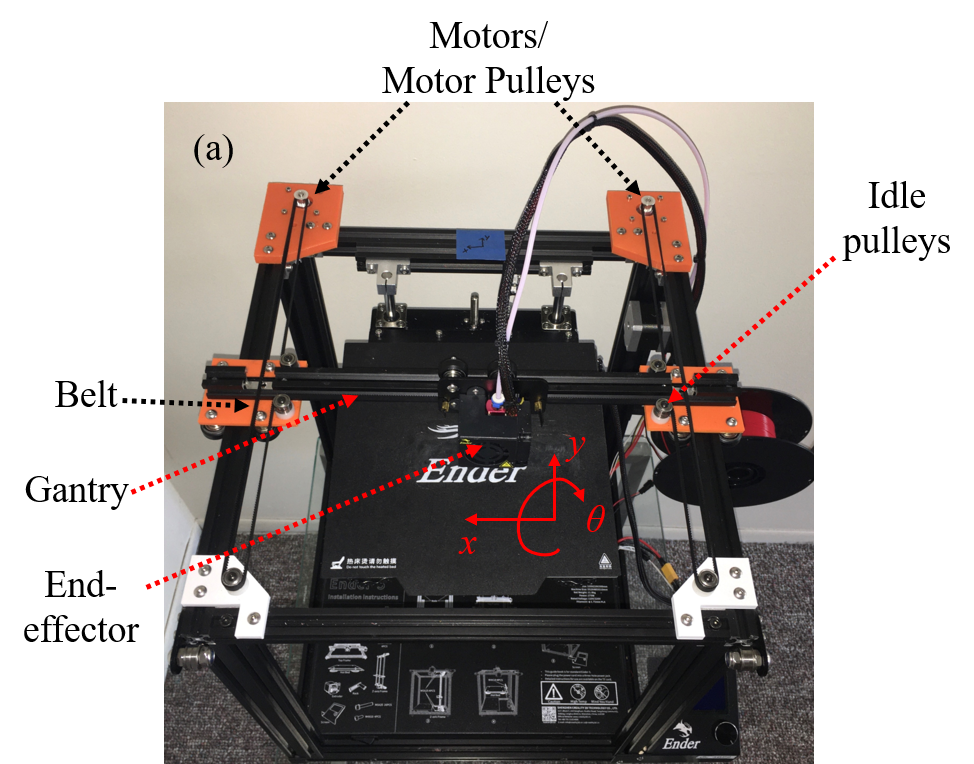}
    \includegraphics[width=.28\textwidth]{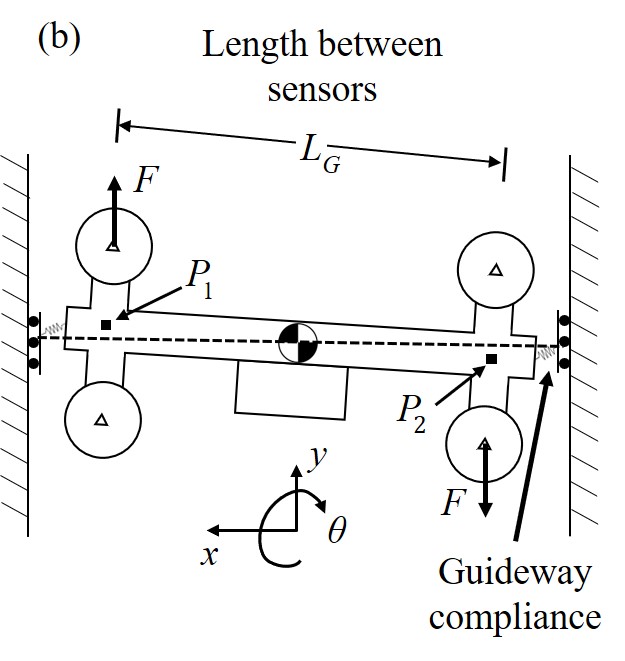}
    \caption{(a) Retrofitted H-frame 3D printer used to validate H-frame dynamic model and conduct experiments in Section IV; (b) Schematic of racking motion due to the force couple F and guideway compliance denoted by a spring.}
    \label{fig:fig03}
\end{figure}

To validate the model of Fig. 2, sine sweep commands at various frequencies were commanded in the $x$ direction of the printer by applying acceleration commands $\ddot{x}_d$ to the stepper motors and measuring $y$-axis accelerations at the locations marked $P_1$ and $P_2$ in Fig. 3(b) with the help of two ADXL335 three-axis accelerometers. The racking angular acceleration is estimated (based on small angle rotations) as
\begin{equation}
    \ddot{\theta} = \frac{\ddot{y}_1 - \ddot{y}_2}{L_G},
\end{equation}
where $\ddot{y}_1$ and $\ddot{y}_2$ are the $y$-axis accelerations measured at $P_1$ and $P_2$ at each end of the bridge (Fig. 3(b)) and $L_G$ is the perpendicular distance between $P_1$ and $P_2$. Accordingly, the frequency response function (FRF) $G_{x\theta}$ is computed using $\ddot{x}_d$ as input and ($\ddot{\theta}$) as output. Figure 4 shows $G_{x\theta}$ determined with the end-effector positioned at $x=0,\pm 30$, and $\pm 60$ mm. The discrepancy between the FRFs is small, supporting the assumption that the gantry location does not significantly influence $G_{x\theta}$. Similarly, $G_{xx}$ and $G_{yy}$ (Fig. 5) are determined by using acceleration commands in the $x$ and $y$ directions, respectively, as inputs and acceleration output measured in the $x$ and $y$ directions with the gantry at $x = 0$ mm. 

\begin{figure}[]
	\centering
    \includegraphics[width=.40\textwidth]{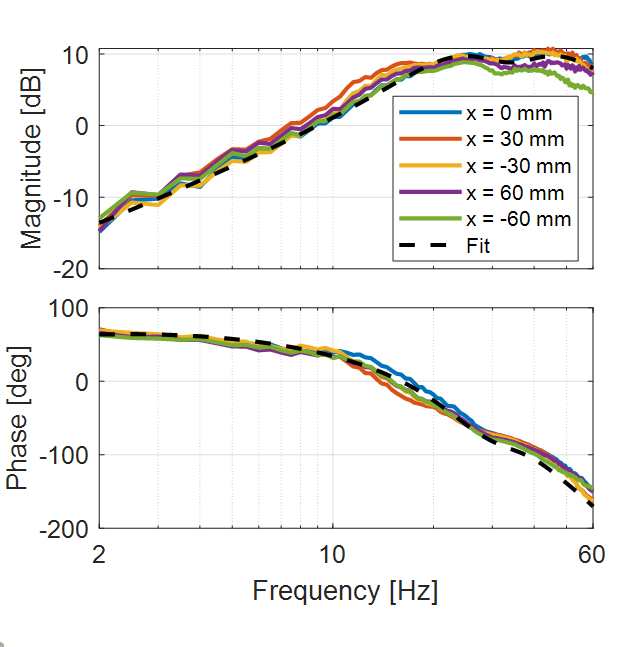}
    \caption{Frequency response functions from $x$-axis command input to $\theta$ output ($G_{x\theta}$) measured with the end-effector positioned at $x=0,\pm 30$, and $\pm 60$ mm. The differences between the FRFs is small. Therefore, they are modeled by a single FRF shown in dashed lines.}
    \label{fig:fig04}
\end{figure}
By curve-fitting the FRFs for $G_{x\theta}$ (Fig. 4), $G_{xx}$ (Fig. 5(a)), and $G_{yy}$ (Fig. 5(b)), discrete transfer functions for each FRF are obtained in the form
\begin{equation}
    \hat{G}(z) = \frac{b_qz^q + b_{q-1}z^{q-1}+\cdots + b_1z+ b_0}{z^d + a_{d-1}z^{d-1} + \cdots + a_1z + a_0}
\end{equation}
where the $\hat{}$ accent is used to denote a model of the actual dynamics, $z$ is the discrete-time forward shift operator, $q$ and $d$ are the degrees of the numerator and denominator polynomials, respectively, and the coefficients of each transfer function are given in Tables I and II. Note that the FRF measured with $x$ at 0 mm is used to fit the transfer function $G_{x\theta}$.
\begin{table*}[]
\centering
\caption{Numerators of fitted transfer functions in equation (4).}
\begin{tabular}{|c|c|c|c|c|c|c|c|c|c|}
\hline
                       & $b_8$ & $b_7$  & $b_6$  & $b_5$   & $b_4$   & $b_3$  & $b_2$  & $b_1$   & $b_0$                   \\ \hline
$\hat{G}_{xx}(z)$      & -     & -      & -      & -       & 0.07632 & -0.231 & -0.236 & -0.0813 & $-2.193\times 10^{-13}$ \\ \hline
$\hat{G}_{x\theta}(z)$ & 4.246 & -29.27 & 87.51  & -147    & 149.9   & -92.76 & 32.26  & -4.865  & $1.808\times 10^{-15}$  \\ \hline
$\hat{G}_{yy}(z)$      & -     & -      & 0.1646 & -0.8951 & 1.975   & -2.199 & 1.231  & -0.2764 & $-9.173\times 10^{-13}$ \\ \hline
\end{tabular}
\end{table*}

\begin{table*}[]
\centering
\caption{Denominators of fitted transfer functions in equation (4).}
\begin{tabular}{|c|c|c|c|c|c|c|c|c|c|}
\hline
                       & $a_8$  & $a_7$ & $a_6$  & $a_5$  & $a_4$  & $a_3$ & $a_2$   & $a_1$                   & $a_0$                   \\ \hline
$\hat{G}_{xx}(z)$      & -      & -     & -      & -      & -3.577 & 4.774 & -2.815  & 0.6175                  & $1.934\times 10^{-33}$  \\ \hline
$\hat{G}_{x\theta}(z)$ & -5.866 & 14.86 & -21    & -17.83 & -9.06  & 2.535 & -0.2987 & $-2.125\times 10^{-17}$ & $2.404\times 10^{-34}$  \\ \hline
$\hat{G}_{yy}(z)$      & -      & -     & -4.813 & 9.386  & -9.344 & 4.845 & -1.14   & 0.06512                 & $-9.037\times 10^{-19}$ \\ \hline
\end{tabular}
\end{table*}

\begin{figure*}[]
	\centering		
	\includegraphics[width=.80\textwidth]{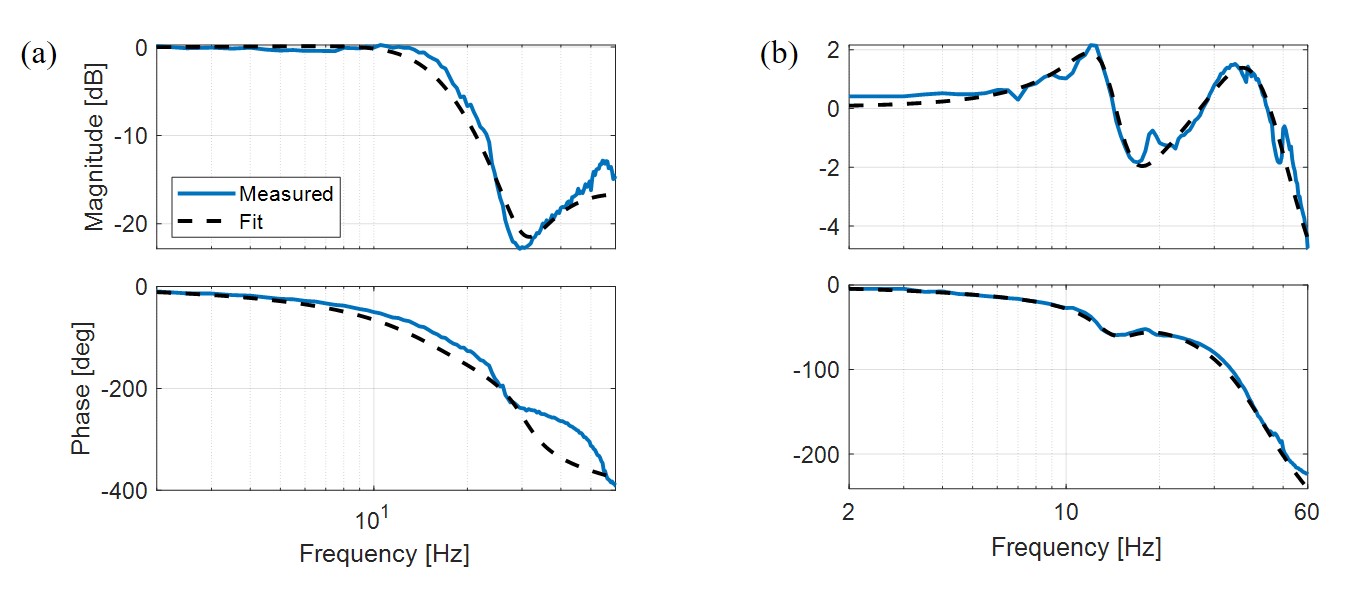}
    \caption{Measured and curve fit FRFs for (a) $G_{xx}$ and (b) $G_{yy}$.}
    \label{fig:fig05}
\end{figure*}

\section{Decoupled LTI and Proposed Coupled LPV Filtered B-splines Approaches}

\subsection{Overview of the decoupled LTI FBS approach (i.e., FBS without racking compensation)}

Figure 6 shows the block diagram of the standard filtered B-splines (FBS) approach for the x-axis of a decoupled multi-axis system, i.e., without racking compensation, as introduced in [38]. It controls a decoupled LTI discrete-time system given by $\mathbf{G}_{xx}$, the lifted system (or matrix) representation of transfer function $G_{xx}$, through a feedforward controller $\mathbf{C}_x$ (see Appendix A for details on the lifted system representation). 

As shown in Fig. 6, let $\mathbf{x}_{d} = [x_{d}(0)\hspace{0.5em} x_{d}(1)\hspace{0.5em} \cdots\hspace{0.5em} x_{d}(E)]^{T}$ represents $E+1$ discrete time steps of the $x$-component of a desired trajectory of a multi-axis machine. Assume that the machine has look-ahead capabilities such that the $E+1$ steps $\mathbf{x}_{d}$ are known in advance. Furthermore, assume that the modified but un-optimized motion command $\mathbf{x}_{dm} = [x_{dm}(0)\hspace{0.5em} x_{dm}(1) \hspace{0.5em} \cdots\hspace{0.5em} x_{dm}(E)]^{T}$ is parameterized using B-splines such that
\begin{equation}
    \begin{bmatrix}
       x_{dm}(0) \\
       x_{dm}(1) \\
       \vdots \\
       x_{dm}(E)
    \end{bmatrix}
    = \underbrace{
    \begin{bmatrix}
       N_{0,m}(\xi_0) & N_{1,m}(\xi_0) & \cdots & N_{n,m}(\xi_0) \\
       N_{0,m}(\xi_1) & N_{1,m}(\xi_1) & \cdots & N_{n,m}(\xi_1) \\
       \vdots & \vdots & \ddots & \vdots \\
       N_{0,m}(\xi_E) & N_{1,m}(\xi_E) & \cdots & N_{n,m}(\xi_E) \\
    \end{bmatrix}
    }_{\mathbf{N}}
    \underbrace{
    \begin{bmatrix}
       p_{x,0} \\
       p_{x,1} \\
       \vdots \\
       p_{x,n}
    \end{bmatrix}
    }_{\mathbf{p}_x}
    \label{eq:fbs_parameterization}
\end{equation}
where $\mathbf{N}$ is the matrix representation of B-spline basis functions of degree $m$, $\mathbf{p}_i$ is a vector of $n+1$ unknown coefficients (or control points), $j=0,1,...,n$, and $\xi \in [0,1]$ is the spline parameter, representing normalized time, which is discretized to $E+1$  uniformly spaced points $\xi_0,\xi_1,...,\xi_E$. The real-valued basis functions, $N_{j,m}(\xi)$, are given by [42]
\begin{equation}
    N_{j,m}(\xi) = \frac{\xi-g_j}{g_{j+m}-g_j}N_{j,m-1}(\xi) + \frac{g_{j+m+1}-\xi}{g_{j+m+1}-g_{j+1}}N_{j+1,m-1}(\xi) \label{eq:bsplines}
\end{equation}
\begin{equation*}
    N_{j,0} =
    \begin{cases}
    1, & g_j \le\xi\le g_{j+1}\\
    0, & \text{otherwise}
    \end{cases}
\end{equation*}
where $\mathbf{g}=[g_0\hspace{0.5em}g_1\hspace{0.5em}\cdots\hspace{0.5em}g_{m+n+1}]^{T}$ is a normalized knot vector defined over $[0,1]$. For convenience, $\mathbf{g}$ is assumed to be uniformly spaced, i.e.,
\begin{equation}
    g_{j} = 
    \begin{cases}
    0, & 0 \le j \le m \\
    \frac{j-m}{n-m+1}, & m+1\le j\le n \\
    1, & n+1 \le j \le m+n+1
    \end{cases}
\end{equation}

\begin{figure*}[]
	\centering		\includegraphics[width=.80\textwidth]{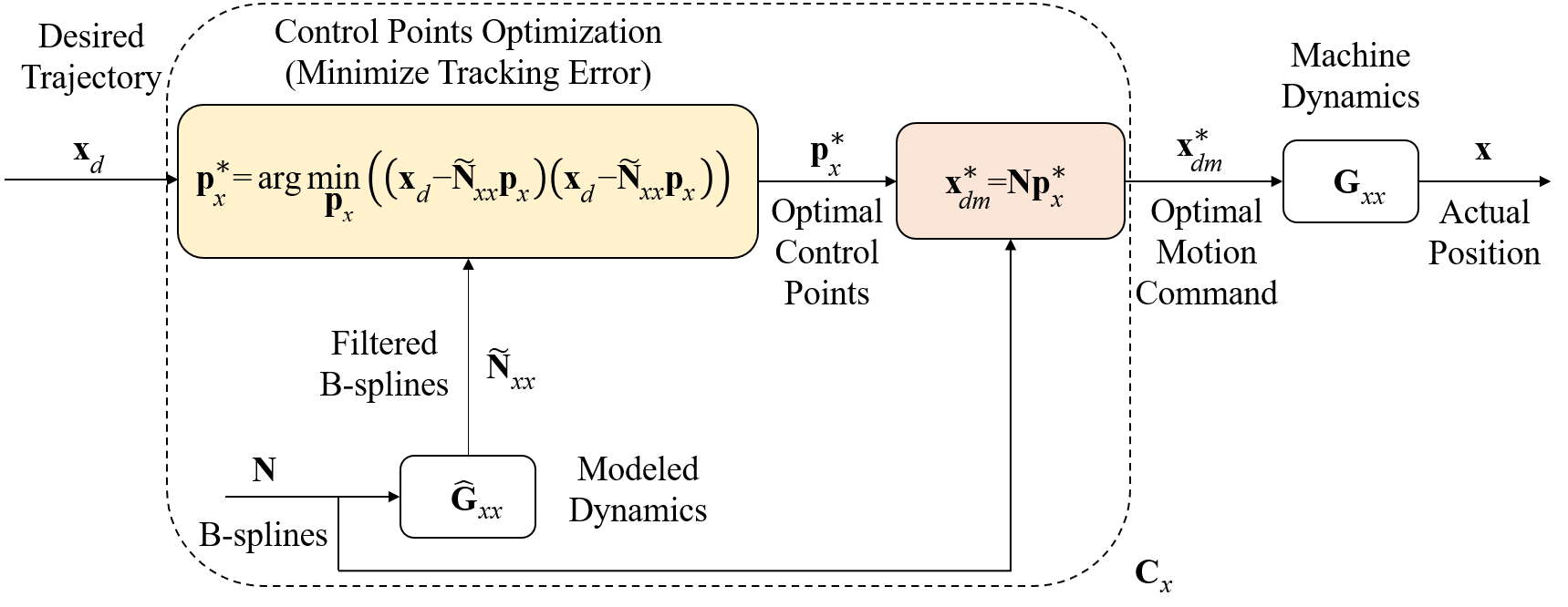}
    \caption{Block diagram of standard FBS method applied to $x$-axis of a decoupled multi-axis system.}
    \label{fig:fig06}
\end{figure*}
Let $\mathbf{x}$ represent the $E+1$ discrete steps of $x$, the motion output of the machine’s $x$-axis. Accordingly, based on the definition of $\mathbf{x}_{dm}$ in Eq. (\ref{eq:fbs_parameterization}), $\mathbf{x}$ can be written as
\begin{equation}
    \mathbf{x} = \tilde{\mathbf{N}}_{xx}\mathbf{p}_{x}
\end{equation}
where $\tilde{\mathbf{N}}_{xx}$ is the filtered B-splines matrix, acquired by filtering each column of $\mathbf{N}$ through $G_{xx}$ (i.e., the matrix product of $\mathbf{N}$ and $\mathbf{G}_{xx}$). Using $\mathbf{x}$ and $\mathbf{x}_d$, the tracking error can be defined as
\begin{equation}
    \mathbf{e}_x = \mathbf{x}_d - \mathbf{x} = \mathbf{x}_d - \tilde{\mathbf{N}}_{xx}\mathbf{p}_x.
\end{equation}
and the optimal control points $\mathbf{p}_{x}^*$ are calculated by minimizing the square of the $L_2$-norm of the tracking error
\begin{align}
    \mathbf{p}_x^* &= \arg\min_{\mathbf{p}_x}(\mathbf{e}_x^{T}\mathbf{e}_x) \\
    & = \arg\min_{\mathbf{p}_x}\Big((\mathbf{x}_d - \tilde{\mathbf{N}}_{xx}\mathbf{p}_x)^{T}(\mathbf{x}_d - \tilde{\mathbf{N}}_{xx}\mathbf{p}_x) \Big)
\end{align}
giving the well-known least squares solution
\begin{equation}
    \mathbf{p}_{x}^* = \Big(\tilde{\mathbf{N}}_{xx}^{T}\tilde{\mathbf{N}}_{xx}\Big)^{-1}\tilde{\mathbf{N}}_{xx}^{T}\mathbf{x}_d = \tilde{\mathbf{N}}_{xx}^{\dagger}\mathbf{x}_{d}
\end{equation}
where the $\dagger$ in the superscript represents the Moore-Penrose inverse (or pseudoinverse) of the matrix. The result can then be used to calculate the optimized motion command $\mathbf{x}_{dm}^* = \mathbf{N}\mathbf{p}_{x}^*$. The same procedure is followed to find the optimal control input for other axes, e.g., $y$-axis.

\noindent \underline{\textit{Remark 3.1}}: The limited-preview version of FBS (LPFBS) [16] relaxes the assumption that $\mathbf{x}_d$ is known in advance and instead uses small windows (batches) of $\mathbf{x}_d$ to achieve on-line implementable control. A brief overview of LPFBS is included in Appendix B.

In regard to H-frame 3D printers, the decoupled LTI implementation of FBS discussed above has two issues due to the introduction of racking. The first is that the motion of the $x$-axis affects the $y$-axis due to racking. Therefore, the $y$-axis cannot be controlled independent of the $x$-axis. The second issue is that the control of the y-axis depends on the position of the end-effector on the $x$-axis. Therefore, a coupled LPV FBS approach is needed to include racking dynamics in H-frame 3D printer control.

\subsection{Proposed coupled LPV FBS approach (i.e., FBS with racking compensation)} 
Noting that the racking model from Section II can be used to predict the error $\Delta y$  from Eqs. (1) and (2), we can use the product of the B-splines matrix $\mathbf{N}$ and $\mathbf{G}_{x\theta}$ to obtain $\tilde{\mathbf{N}}_{x\theta}$. Therefore, using Eqs. (5) and (8), we have 
\begin{equation}
    \mathbf{\theta} = \tilde{\mathbf{N}}_{x\theta}\mathbf{p}_{x}
\end{equation}
and
\begin{equation}
    \Delta\mathbf{y} = \mathbf{D}_{x_d}\tilde{\mathbf{N}}_{x\theta}\mathbf{p}_{x},
\end{equation}
where $\mathbf{D}_{x_d} = diag(\mathbf{x}_d)$. The tracking error for each axis can then be expressed as 
\begin{align}
    \mathbf{e}_x & = \mathbf{x}_d - \mathbf{x} = \mathbf{x}_d - \tilde{\mathbf{N}}_{xx}\mathbf{p}_x \\
    \mathbf{e}_y & = \mathbf{y}_d - \mathbf{y} = \mathbf{y}_d - (\mathbf{D}_{x_d}\tilde{\mathbf{N}}_{x\theta}\mathbf{p}_{x} + \tilde{\mathbf{N}}_{yy}\mathbf{p}_y)
\end{align}
and the optimal control points can be calculated to minimize the squared $L_2$-norm of sum of the tracking errors
\begin{equation}
    \mathbf{p}^* = \arg\min_{\mathbf{p}}(\mathbf{e}^{T}\mathbf{e})
\end{equation}
where $\mathbf{e}=[\mathbf{e}_x\hspace{0.5em}\mathbf{e}_y]^{T}$, which gives
\begin{equation}
    \mathbf{p}^* =
    \begin{bmatrix}
       \mathbf{p}_{x}^* \\
       \mathbf{p}_{y}^*
    \end{bmatrix}
    =
    \begin{bmatrix}
       \tilde{\mathbf{N}}_{xx} & \mathbf{0} \\
       \mathbf{D}_{x_d}\tilde{\mathbf{N}}_{x\theta} & \tilde{\mathbf{N}}_{yy}
    \end{bmatrix}^\dagger
    \begin{bmatrix}
       \mathbf{x}_d \\
       \mathbf{y}_d
    \end{bmatrix}.
\end{equation}
The formulation of the coupled LPV FBS controller in Eq. (18) can be computationally cumbersome during on-line implementation due to the size of the matrix that is inverted. Furthermore, when using LPFBS, since the H-frame is an LPV system, we cannot pre-invert the matrix as is done with LTI FBS, to reduce the computational load (see Appendix B). Therefore, we decouple the matrices to eliminate the need to compute the pseudoinverse of large matrices during implementation. The control points are thus determined sequentially: first $\mathbf{p}_{x}^*$ using Eq. (12), then $\mathbf{p}_{y}^*$ by rearranging Eq. (16) and considering $\mathbf{p}_{x}^*$ as a known input
\begin{equation}
    \mathbf{p}_{y}^* = \tilde{\mathbf{N}}_{yy}^\dagger(\mathbf{y}_d - \mathbf{D}_{x_d}\tilde{\mathbf{N}}_{x\theta}\mathbf{p}_{x}^*).
\end{equation}
Note that in the decoupled approximation using Eqs. (12) and (19), we are inverting the same matrices from LTI FBS, and can pre-invert the matrices for on-line implementation. In Section III.C below, we consider the effects of this decoupled approximation on the tracking accuracy and computational complexity of the proposed controller.

\noindent\underline{\textit{Remark 3.2}}: The LPFBS form of the proposed FBS controller with racking compensation is discussed in Appendix C.

\subsection{Tracking accuracy and computational complexity of coupled and decoupled LPV controllers}

The system with racking can be expressed in the lifted system representation (see Appendix A) as
\begin{equation}
    \mathbf{G} = 
    \begin{bmatrix}
       \mathbf{G}_{xx} & \mathbf{0} \\
       \mathbf{G}_{xy} & \mathbf{G}_{yy}
    \end{bmatrix}
\end{equation}
where $\mathbf{G}_{xy}=\mathbf{D}_{x_d}\mathbf{G}_{x\theta}$. The inverse of $\mathbf{G}$ is given by
\begin{equation}
    \mathbf{G}^{-1} = 
    \begin{bmatrix}
       \mathbf{G}_{xx}^{-1} & \mathbf{0} \\
       -\mathbf{G}_{yy}^{-1}\mathbf{G}_{xy}\mathbf{G}_{xx}^{-1} & \mathbf{G}_{yy}^{-1}
    \end{bmatrix}
\end{equation}
and the optimal control inputs $\mathbf{x}_{dm}^*$ and $\mathbf{y}_{dm}^*$ are given by
\begin{align}
    \mathbf{x}_{dm}^* & = \mathbf{G}_{xx}^{-1}\mathbf{x}_{d} \\
    \mathbf{y}_{dm}^* & = -\mathbf{G}_{yy}^{-1}\mathbf{G}_{xy}\mathbf{G}_{xx}^{-1}\mathbf{x}_{d} + \mathbf{G}_{yy}^{-1}\mathbf{y}_{d} 
\end{align}
It can be shown (see [38]) that for $n=E$, $\tilde{\mathbf{N}}_{xx}=\mathbf{G}_{xx}$, $\tilde{\mathbf{N}}_{yy}=\mathbf{G}_{yy}$ and the pseudoinversion in Eqs. (12) and (19) become matrix inversion of $\mathbf{G}_{xx}$ and $\mathbf{G}_{yy}$, respectively. Therefore, for $n=E$, the motion command for the decoupled controller, proposed in Section III.B can be expressed as
\begin{align}
    \mathbf{x}_{dm}^* & = \mathbf{G}_{xx}^{-1}\mathbf{x}_{d} \\
    \mathbf{y}_{dm}^* & = \mathbf{G}_{yy}^{-1}(\mathbf{y}_{d} - \mathbf{G}_{xy}\mathbf{G}_{xx}^{-1}\mathbf{x}_{d}) = \mathbf{G}_{yy}^{-1}\mathbf{y}_{d} - \mathbf{G}_{yy}^{-1}\mathbf{G}_{xy}\mathbf{G}_{xx}^{-1}\mathbf{x}_{d}
\end{align}
Note that Eqs. (23) and (25) are identical which shows that the decoupled LPV FBS approach is exactly the same as the inversion of the coupled LPV system when $n=E$. When $n<E$, it approximates the coupled LPV system using the FBS approach. Equations (12) and (19) are, therefore, another way of approximating the coupled LPV system using the FBS approach.

The computational complexity of the Moore-Penrose inverse, computed using singular value decomposition, is given by $O(lv^2)$ where $l$ and $v$ are the number of rows and columns, respectively, of the matrix to be inverted [43]. We note that the size of the coupled LPV FBS matrix from the Section 3.2 $2(E+1) \times (n_x + n_y + 2)$, where we consider the number of basis functions in the $x$ and $y$ axes independently, and the size of the decoupled matrices are $(E+1)\times (n_x+1)$ and $(E+1)\times (n_y+1)$. Assuming $n = n_x = n_y = E$, the computational complexity of the coupled and decoupled LPV FBS approaches are 
\begin{align}
    O_{c}\Big((2E)(2n)^2\Big) & = O_{c}\Big((2n)^3\Big) = O_{c}(8n^3), \\
    O_{d}(En^2+En^2) & = O_{d}\Big(E(n^2 + n^2)\Big) = O_{d}(En^2) = O_{d}(n^3),
\end{align}
where $O_{c}$ and $O_{d}$ are the computational orders of the coupled and decoupled approximations, respectively. The expressions in Eqs. (26) and (27) indicate that the decoupled matrix approximation has much lower computational complexity. The implications of using the decoupled or coupled approximations of the LPV dynamics on racking compensation accuracy and computation time are explored further via simulations in the following section.

\section{Simulations and Experiments}

\subsection{Simulations}
Figure 7 shows the 120-by-20 mm rectangular motion path used for our simulations. The trajectory along the path was generated using a jerk-limited motion profile [44] with a maximum velocity, acceleration, and jerk of 150 mm/s, $1\times 10^{4}$ mm/s$^2$, and $5\times 10^{7}$ mm/s$^3$, respectively, and was sampled at $T_s = 1$ ms, leading to $E+1 = 1944$ trajectory points. The model derived and presented in Section II, was used as the dynamics to simulate the time response of the H-frame in MATLAB. 

\begin{figure}[]
	\centering
    \includegraphics[width=.48\textwidth]{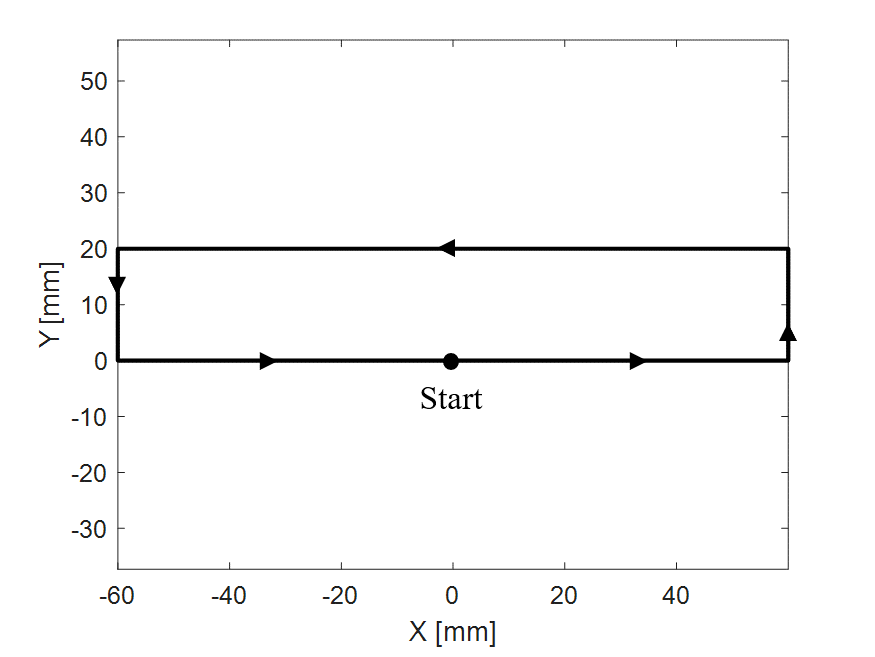}
    \caption{Rectangular path (with 120 mm length and 20 mm width) used to simulate the time response of the H-frame 3D Printer. The motion command starts at $\{0,0\}$ and traverses the rectangle in the counterclockwise direction as indicated by the arrows.}
    \label{fig:fig07}
\end{figure}
\subsubsection{Comparison of decoupled and coupled FBS approaches with racking compensation}
In Fig. 8(a), we compare the tracking accuracy of the decoupled and coupled FBS approximations from Section III.B when $n<E$ by recording the root-mean-squared (RMS) contour (i.e., path deviation) error of the output for each method. The number of basis functions was selected to span fractional values of $E$, namely $n = [0.01, 0.05, 0.1, 0.15, 0.2, 0.25]E$ (rounded to the nearest integer), the B-spline degree was $m=5$, and the knot vector is defined as in Section III.A. Following from the analysis in Section III.C, note that the RMS contour error between the two methods is similar for all the $n$ values, indicating that using the decoupled approach yields similar tracking performance to the coupled approach. Similarly, we validated the computational complexity analysis from Section III.C by comparing the computation time between the two approaches in Fig. 8(b). Note that the coupled approach requires significantly more computation time as $n$ increases. Based on the analysis in Section III.C and the results shown in Fig. 8, the decoupled implementation of the coupled LPV FBS controller (i.e., FBS with racking compensation) will be used in all simulations and experiments hereinafter.
\begin{figure*}[]
	\centering
    \includegraphics[width=.80\textwidth]{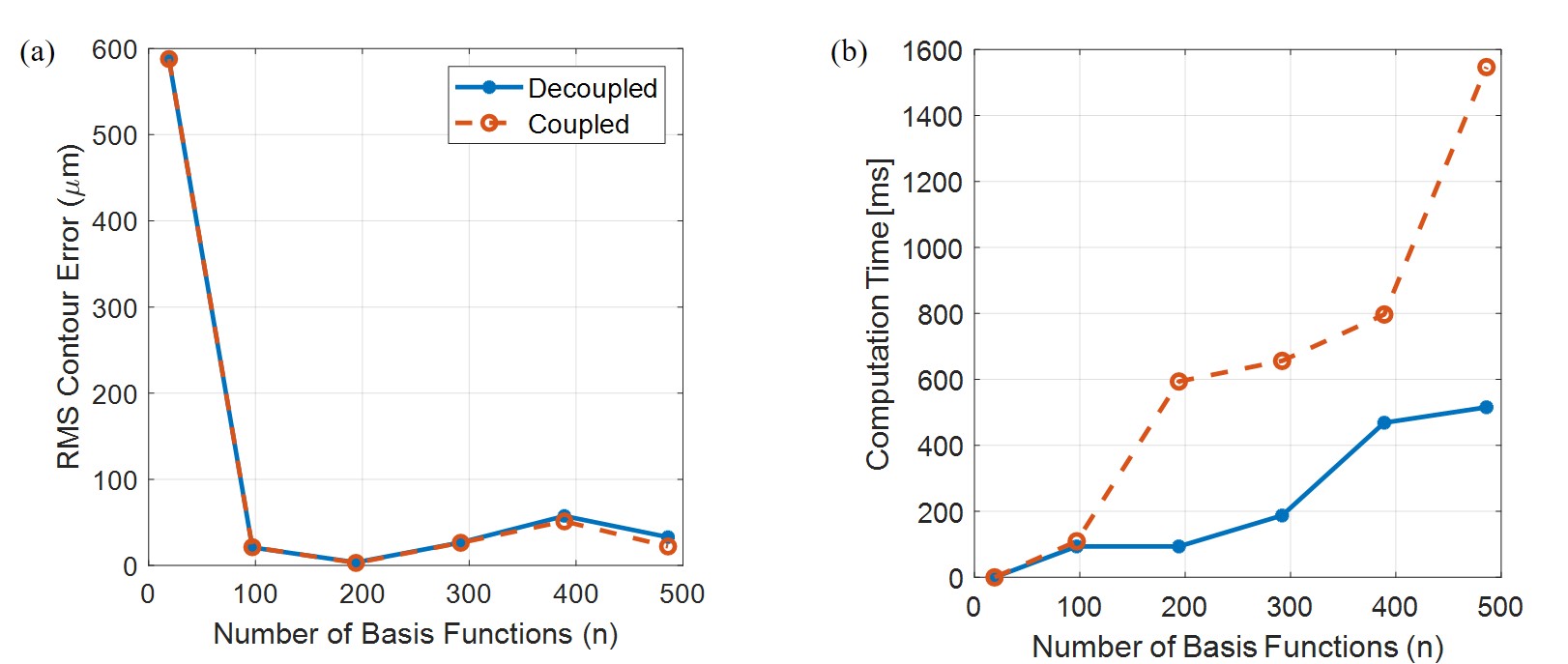}
    \caption{(a) RMS contour errors and (b) computation times for the simulated time response of the decoupled (solid line) and coupled (dashed line) LPV FBS H-frame controllers as a function of the number of basis functions used to parameterize the trajectory.}
    \label{fig:fig08}
\end{figure*}

\subsubsection{Comparison of FBS controller with and without racking compensation}
The simulated response of the H-frame machine (using the same trajectory from Fig. 7), controlled with $n=125$ B-spline basis functions, is shown in Fig. 9. Using the FBS controller without racking compensation, the racking errors can be seen at the corners of the rectangle as it is traversed (Fig. 9(a)). These errors are mitigated using the proposed FBS controller with racking compensation and can be seen in Fig. 9(b), where the contour errors are shown as a function of the path length (i.e., perimeter) of the rectangle. The RMS contour error across the trajectory for the standard FBS controller was 117 $\mu$m, compared to 9 $\mu$m for the FBS controller with racking compensation—a 13 times improvement.

\begin{figure}[]
	\centering
    \includegraphics[width=.48\textwidth]{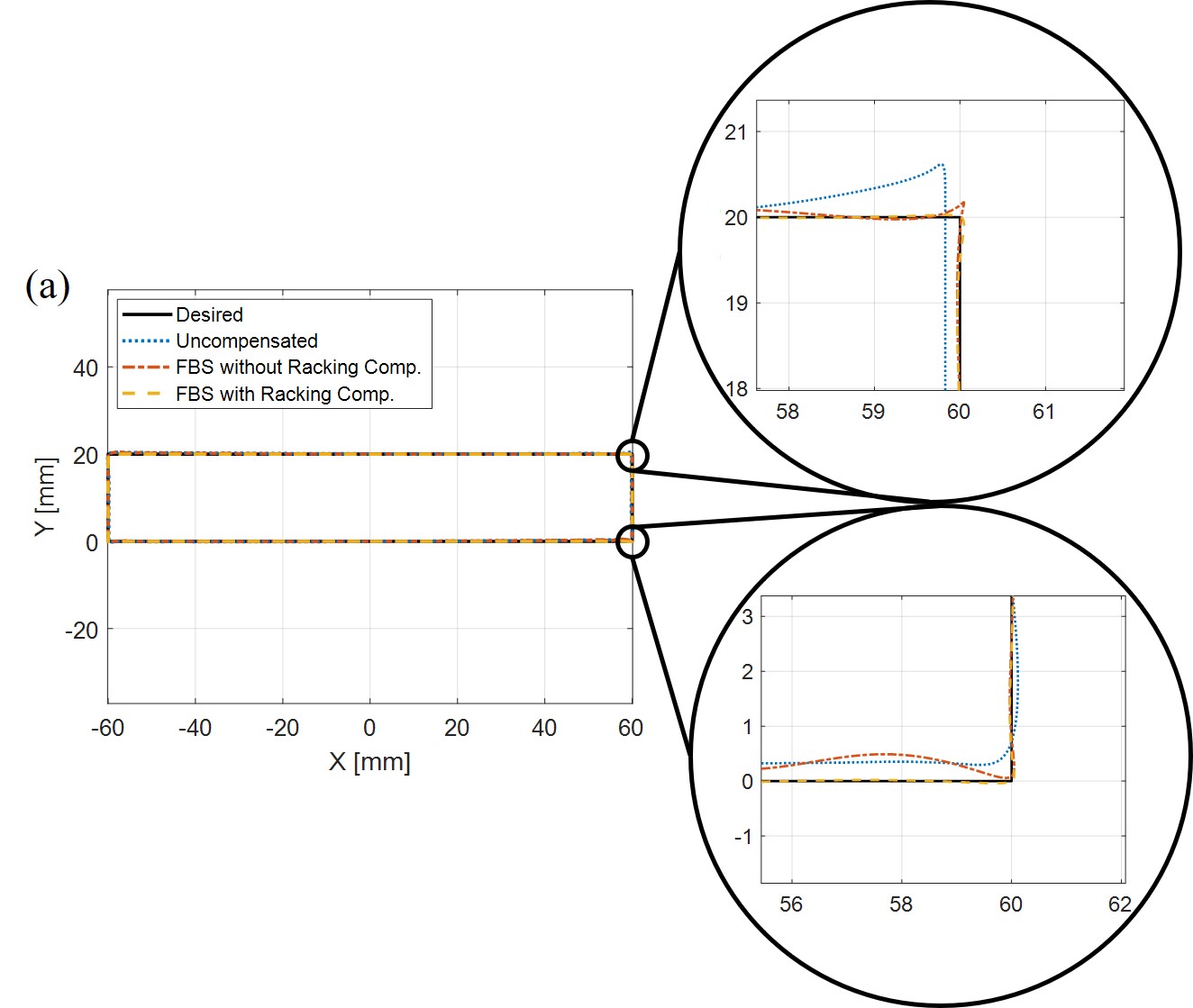}
    \includegraphics[width=.48\textwidth]{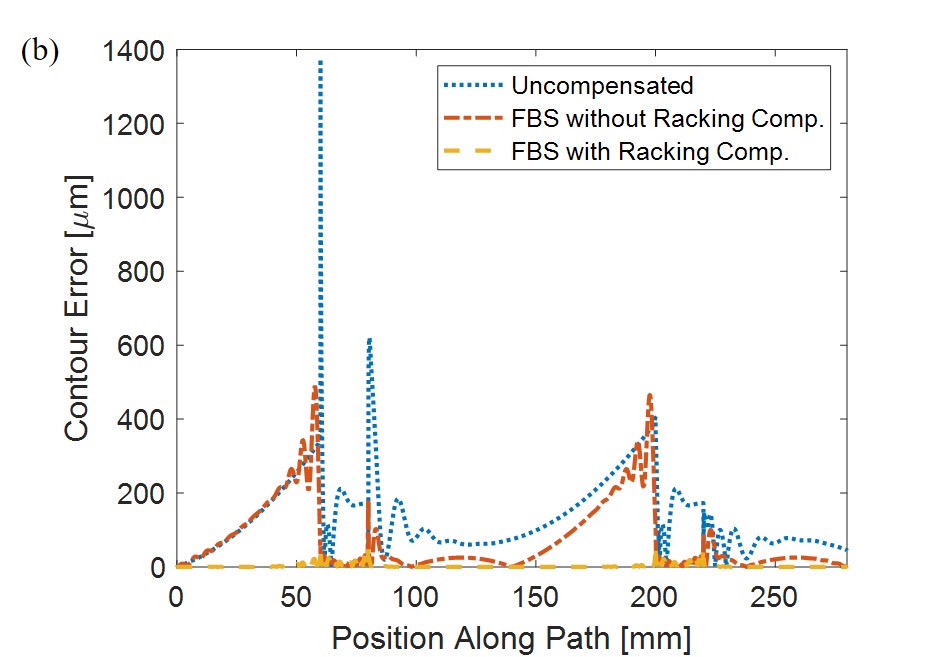}
    \caption{(a) Simulated time response of the rectangular trajectory for the uncompensated trajectory (dotted line) as well as the trajectories generated using FBS controllers without racking compensation (dot-dash line) and with racking compensation (dashed line); (b) Contour error of the simulated time response of each control scheme as a function of position along path (i.e., perimeter of the rectangle).}
    \label{fig:fig09}
\end{figure}

\subsection{Experiments}
The same rectangle profile was extruded to a height of 10 mm and printed on the H-frame 3D printer. The G-code for the trajectory was generated using the open-source Ultimaker Cura® software with a wall speed of 150 mm/s (which is the same as used in Section IV.A) and an infill speed of 60 mm/s. The acceleration and jerk limits used for trajectory generation were also the same as those used in Section IV.A. The wall thickness and layer height were selected to be 0.8 mm and 0.1 mm, respectively. To ensure adhesion to the bed, the first four layers (up to 0.4 mm) were printed at a speed of 20 mm/s. The CAD model of the rectangular prism can be seen in Figure 10(a) along with example parts printed without any compensation (Figure 10(b)), printed using FBS without racking compensation (Figure 10(c)), and printed using FBS with racking compensation (Figure 10(d)).
\begin{figure}[]
	\centering
    \includegraphics[width=.48\textwidth]{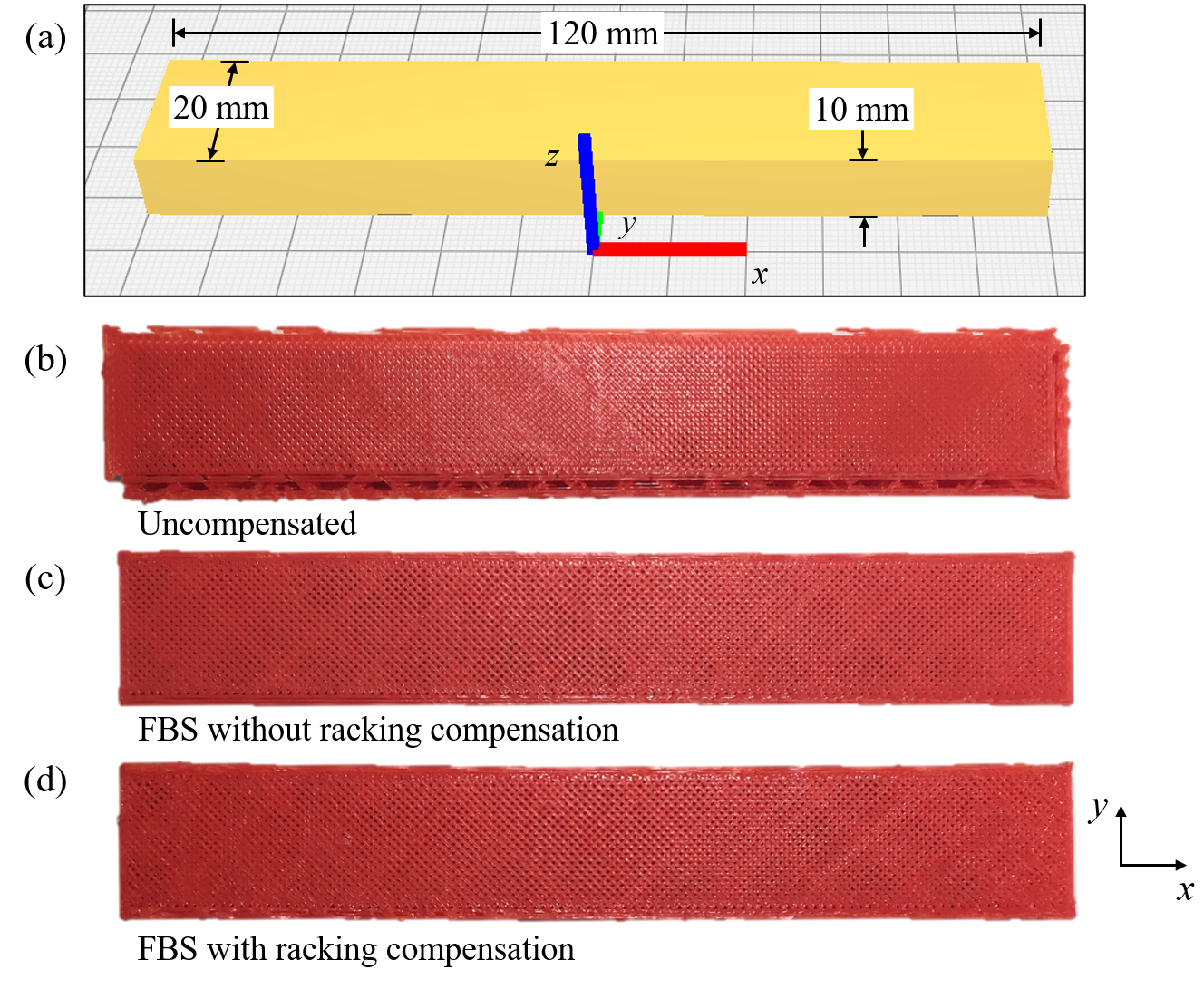}
    \caption{(a) CAD model of the part in Ultimaker Cura® and examples of parts printed (b) without any compensation, (c) using FBS without racking compensation, and (d) using FBS with racking compensation.}
    \label{fig:fig10}
\end{figure}
We printed 15 copies of the model using three different compensation strategies (5 copies for each): (1) no compensation, (2) FBS without racking compensation, and (3) the proposed FBS with racking compensation. The limited-preview version of FBS, i.e., LPFBS, was used in both FBS cases, with parameters $n_{up}=11$, $n_{C} = 22$, $L_{C}=220$, $m=5$. and $L=20$ (see Appendix B and C for more details). As shown in Fig. 10(b), the parts printed without compensation suffered from printing layer shifts during the printing process, like the layer shifts the authors observed in [16]. The FBS approach (with and without racking compensation) provides sufficient compensation to remove the layer shifts from the printed part. Nevertheless, the width, $w$, of each part was measured at the left, middle, and right side using Husky digital calipers (model\# 1467H, 10 $\mu$m resolution) and compared to the desired width of $w_{d} = 20$ mm. Figure 11 shows a box-and-whisker plot of the absolute value of the width error ($\Delta w = |w_d - w|$) overlaid with the RMS width error, $\Delta w_{rms}$. The proposed FBS controller with racking compensation improves the median $\Delta w$ by 61\% and 78\% compared to FBS without racking compensation and no compensation, respectively. The proposed controller also improves $\Delta w_{rms}$ by 43\% and 68\% compared to FBS without racking compensation and no compensation, respectively.
\begin{figure}[]
	\centering
    \includegraphics[width=.48\textwidth]{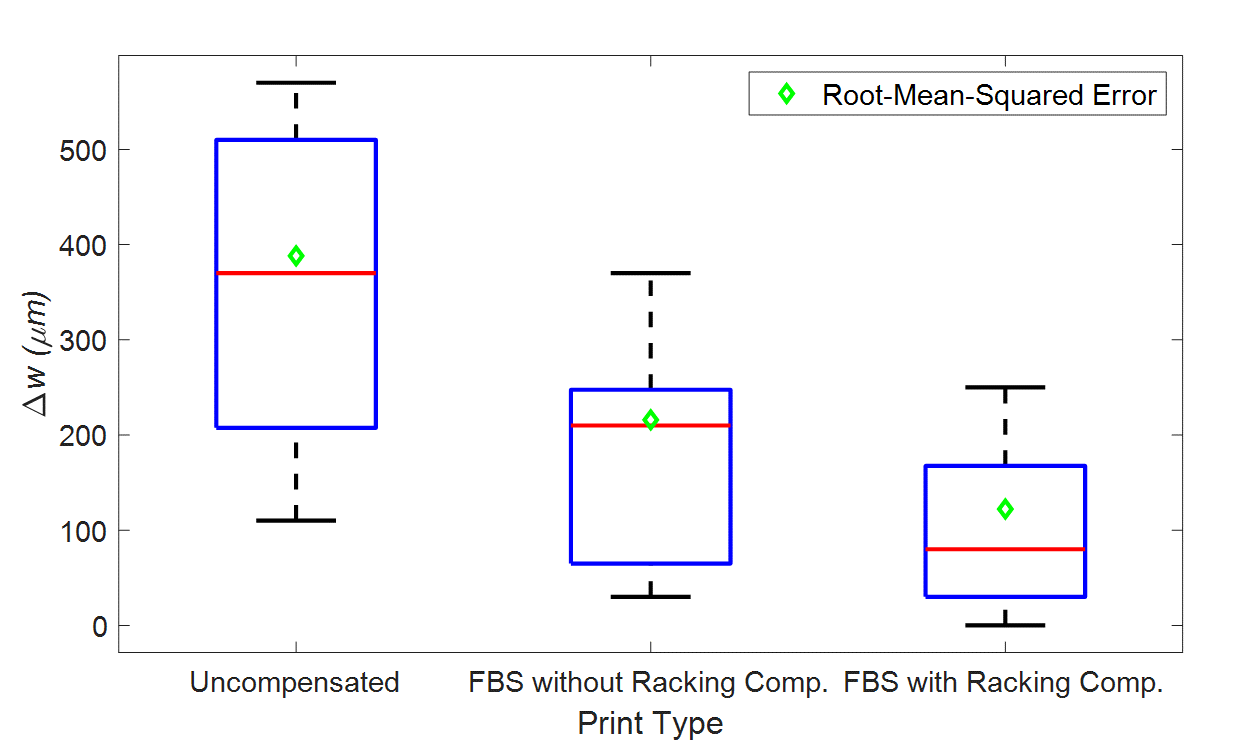}
    \caption{Box-and-whisker plot of measured absolute width error, $\Delta w$, of printed parts compared to the desired width of 20 mm for each of the compensation strategies: (1) no compensation, (2) FBS without racking compensation, and (3) FBS with racking compensation. The red horizontal lines represent the median of $\Delta w$, which is overlaid with root-mean-square width error, $\Delta w_{rms}$ (diamonds).}
    \label{fig:fig11}
\end{figure}

\section{Conclusions}

H-frame 3D printer architectures hold potential to achieve higher speeds and improved dynamic performance compared to traditional serial stack 3D printers due to their use of stationary motors. However, these benefits come at the cost of racking errors, caused by parasitic torsional motions, which limit their static and dynamic accuracy. This paper proposes a purely software-based approach for compensating racking errors on H-frame 3D printers using the filtered B-spline (FBS) feedforward controller that has been used to improve the performance of 3D printers in prior work [16,18,19]. However, to compensate racking errors, the proposed FBS controller is designed to address coupled linear parameter varying dynamics rather than decoupled linear time invariant dynamics addressed in prior work.  A decoupled approximation of the proposed coupled FBS controller, that significantly reduces computational complexity with little or no sacrifice to error compensation accuracy, is developed and validated analytically and numerically. The decoupled FBS controller with racking compensation is benchmarked in simulation and experiments on an H-frame 3D printer against the standard FBS controller without racking compensation. Using the proposed approach, 13 times and 43\% improvement in the shape accuracy of a printed part is observed in simulations and experiments, respectively. 
A major practical benefit of the proposed software-based approach for racking error compensation is that it reduces racking errors without requiring mechanical modification of a 3D printer. Hence it can be applied to existing H-frame 3D printers. It can also be used to augment other mechanical or software-based approaches for addressing racking errors, like the use of stiffer guideways, dampers and feedback controllers. This paper shows the potential of software-based compensation approaches to improve the dynamic performance of parallel-axis motion-stage architectures (often used on 3D printers). Future work will explore the application of the FBS approach to other parallel-axis motion-stage architectures that suffer from coupled dynamics and parasitic errors, like Delta 3D printers.




\section*{APPENDIX}

\subsection{Lifted system representation of a digital filter}

Consider filter $p$, input signal $u$, and output signal $y$ defined as:
\begin{align}
    p = & \{p_{-2}\hspace{0.5em}p_{-1}\hspace{0.5em}p_{0}\hspace{0.5em}p_{1}\hspace{0.5em}p_{2}\} \label{eq:lsr_filter} \\
    u = & \{u_{0}\hspace{0.5em} u_{1}\hspace{0.5em}u_{2}\} \\
    y = & \{y_{0}\hspace{0.5em} y_{1}\hspace{0.5em} y{1}\} \label{eq:lsr_output}
\end{align}
Signals $y$ and $u$ and filter $p$ are related by the convolution operator as follows:
\begin{equation}
    y = u*p
    \label{eq:convolution}
\end{equation}
From Eqs. (\ref{eq:lsr_filter})-(\ref{eq:convolution}),
\begin{align}
    y_{0} = & p_{0}u_{0} + p_{-1}u_{1} + p_{-2}u_{2} \\
    y_{1} = & p_{1}u_{0} + p_{0}u_{1} + p_{-1}u_{2} \\
    y_{2} = & p_{2}u_{0} + p_{1}u_{1} + p_{0}u_{2}
\end{align}
This can be expressed in matrix form as
\begin{equation}
    \begin{bmatrix}
       y_{0} \\
       y_{1} \\
       y_{2}
    \end{bmatrix}
    =
    \begin{bmatrix}
       p_{0} & p_{-1} & p_{-2} \\
       p_{1} & p_{0} & p_{-1} \\
       p_{2} & p_{1} & p_{0}
    \end{bmatrix}
    \begin{bmatrix}
       u_{0} \\
       u_{1} \\
       u_{2}
    \end{bmatrix}
    \label{eq:lsr_matrix}
\end{equation}
Note that the main diagonal element ($p_{0}$) represents the influence of the current input on the current output; the first upper diagonal element ($p_{-1}$) represents the influence of the succeeding input on the current output and the second upper diagonal element ($p_{-2}$)  represents the influence of the second succeeding input on the current output. Similarly, the first ($p_{1}$) and second lower ($p_{2}$) elements represent the influence of the first and second preceding inputs on the current output, respectively. Hence, the discrete time (or $z$) transform of $p$ obtained from Eq. (\ref{eq:lsr_matrix}) is given by
\begin{equation}
    p_{2}z^{-2} + p_{1}z^{-1} + p_{0}z^{0} + p_{-1}z^{1} + p_{-2}z^{2}
\end{equation}
which is in accordance with the time-domain definition given in Eqs. (\ref{eq:lsr_filter})-(\ref{eq:lsr_output}).

\subsection{The limited-preview filtered B-splines approach}
The limited-preview FBS approach [16] aims to generate optimal feedforward control inputs in sequential windows (batches) of the desired trajectory $\mathbf{x}_d$. Since $\mathbf{x}_d$ is not assumed to be fully known a priori, an un-normalized and open-ended knot vector is used and defined as
\begin{equation}
    \bar{g}_{j} = 
    \begin{cases}
    0, & 0 \le j \le m \\
    (j-m)LT_{s}, & j \le m+1
    \end{cases}
\end{equation}
where $j$ and $m$ are as defined in Sec. III, and $L\ge 1$ represents the uniform spacing of the knot vector elements as an integer multiple of the sampling time $T_{s}$. With the un-normalized knot vector, $N_{j,m}$ is expressed as a function of time $t$ by replacing $\xi$ with $t$ and $g_j$ with $\bar{g}_j$ in Eq. (\ref{eq:bsplines}), and the function is sampled at $t_{k_n}=k_nT_{s}$ to formulate $\mathbf{N}$ as in Eq. (\ref{eq:fbs_parameterization}). The tracking problem is solved in batches as
\begin{multline}
    \bar{\mathbf{e}} = \mathbf{x}_d - \bar{\mathbf{N}}\bar{\mathbf{p}} \Leftrightarrow \\
    \begin{bmatrix}
       \bar{\mathbf{e}}_{\text{P}} \\
       \bar{\mathbf{e}}_{\text{C}} \\
       \bar{\mathbf{e}}_{\text{F}} 
    \end{bmatrix}
    =
    \begin{bmatrix}
       \mathbf{x}_{d,\text{P}} \\
       \mathbf{x}_{d,\text{C}} \\
       \mathbf{x}_{d,\text{F}} 
    \end{bmatrix}
    -
    \begin{bmatrix}
       \bar{\mathbf{N}}_{\text{P}} & \mathbf{0} & \mathbf{0} \\
       \bar{\mathbf{N}}_{\text{PC}} & \bar{\mathbf{N}}_{\text{C}} & \mathbf{0} \\
       \mathbf{0} & \bar{\mathbf{N}}_{\text{CF}} & \bar{\mathbf{N}}_{\text{F}}
    \end{bmatrix}
    \begin{bmatrix}
       \bar{\mathbf{p}}_{\text{P}} \\
       \bar{\mathbf{p}}_{\text{C}} \\
       \bar{\mathbf{p}}_{\text{F}} 
    \end{bmatrix}
\end{multline}
where subscripts P, C, and F denote the past, current, and future batches, respectively, and the bar on the matrices and vectors indicates that the impulse response of the transfer function used for filtering the B-splines is truncated. Using local least squares, the optimal coefficients of the current batch can be computed as 
\begin{equation}
    \bar{\mathbf{p}}_{\text{C}}^{*} = (\bar{\mathbf{N}}_{\text{C}}^{T}\bar{\mathbf{N}}_{\text{C}})^{-1}\bar{\mathbf{N}}_{\text{C}}\Big(\mathbf{x}_{d,\text{C}} - \bar{\mathbf{N}}_{\text{PC}}\bar{\mathbf{p}}_{\text{P}}\Big)
\end{equation}
where $\bar{\mathbf{p}}_{\text{P}}$ denotes the coefficients calculated in the last batch. Note that the information from the future batch is not considered while calculating coefficients for the current batch. Also note that the matrix $\bar{\mathbf{N}}_{\text{C}}$ can be pre-inverted for each region, stored in a look-up table, and applied for each corresponding batch.

\begin{figure}[]
	\centering		
	\includegraphics[width=.40\textwidth]{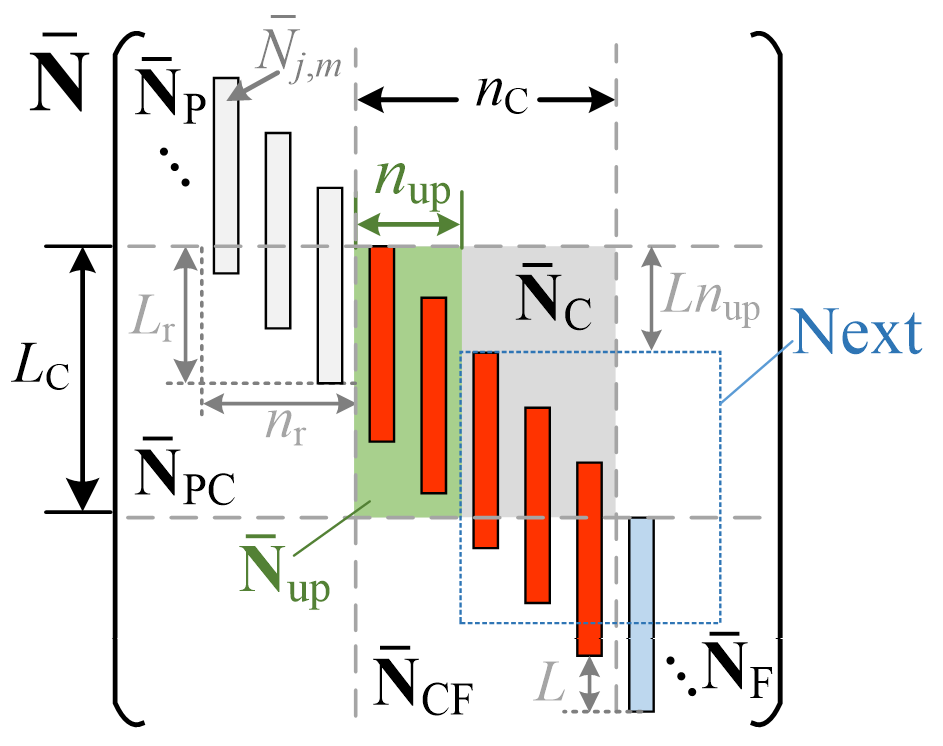}
    \caption{Illustration of the limited-preview filtered B-splines (LPFBS) approach [16].}
    \label{fig:imgB1}
\end{figure}

The dimensions of the current window are defined by $L_C$ and $n_C$, where $L_C$ is the number of trajectory points considered in the current batch and $n_C$ is the number of B-spline coefficients. Note that although $n_C$ coefficents are computed, only $n_{up}$ are updated in each window (see Fig. \ref{fig:imgB1}).

\subsection{Limited-preview filtered B-splines for coupled LPV controller}

From Eqs. (19) and (38), a natural extension of LPFBS can be made for the coupled LPV system. The tracking error is given by

\begin{multline}
    \bar{\mathbf{e}} = \mathbf{r}_d - \bar{\mathbf{N}}_{r}\bar{\mathbf{p}}_{r} \Leftrightarrow \\
    \begin{bmatrix}
       \bar{\mathbf{e}}_{r,\text{P}} \\
       \bar{\mathbf{e}}_{r,\text{C}} \\
       \bar{\mathbf{e}}_{r,\text{F}} 
    \end{bmatrix}
    =
    \begin{bmatrix}
       \mathbf{r}_{d,\text{P}} \\
       \mathbf{r}_{d,\text{C}} \\
       \mathbf{r}_{d,\text{F}} 
    \end{bmatrix}
    -
    \begin{bmatrix}
       \bar{\mathbf{N}}_{r,\text{P}} & \mathbf{0} & \mathbf{0} \\
       \bar{\mathbf{N}}_{r,\text{PC}} & \bar{\mathbf{N}}_{r,\text{C}} & \mathbf{0} \\
       \mathbf{0} & \bar{\mathbf{N}}_{r,\text{CF}} & \bar{\mathbf{N}}_{r,\text{F}}
    \end{bmatrix}
    \begin{bmatrix}
       \bar{\mathbf{p}}_{r,\text{P}} \\
       \bar{\mathbf{p}}_{r,\text{C}} \\
       \bar{\mathbf{p}}_{r,\text{F}} 
    \end{bmatrix}
\end{multline}
where $\mathbf{r}_d = [\mathbf{x}_d \hspace{0.5em} \mathbf{y}_d]^{T}$, and the subscript $r$ denotes matrices and vectors related to  $\mathbf{r}_d$. Expanding the tracking error of the current batch as
\begin{multline}
    \bar{\mathbf{e}}_{r,\text{C}} =
    \begin{bmatrix}
       \mathbf{x}_{d,\text{C}} \\
       \mathbf{y}_{d,\text{C}}
    \end{bmatrix} 
    - \\
    \begin{bmatrix}
       \bar{\mathbf{N}}_{x,\text{PC}} & \mathbf{0} & \bar{\mathbf{N}}_{x,\text{C}} & \mathbf{0} \\
       \mathbf{D}_{x_{d,\text{PC}}}\bar{\mathbf{N}}_{x\theta,\text{PC}} & \bar{\mathbf{N}}_{y,\text{PC}} & \mathbf{D}_{\mathbf{x}_{d,\text{C}}}\bar{\mathbf{N}}_{x\theta,\text{C}} & \bar{\mathbf{N}}_{y,\text{C}}
    \end{bmatrix}
    \begin{bmatrix}
       \bar{\mathbf{p}}_{x,\text{P}} \\
       \bar{\mathbf{p}}_{y,\text{P}} \\
       \bar{\mathbf{p}}_{x,\text{C}} \\
       \bar{\mathbf{p}}_{y,\text{C}} 
    \end{bmatrix}
\end{multline}
the coefficients for the current batch are calculated as
\begin{multline}
    \begin{bmatrix}
       \bar{\mathbf{p}}_{x,\text{C}} \\
       \bar{\mathbf{p}}_{y,\text{C}}
    \end{bmatrix}
    =
    \begin{bmatrix}
       \bar{\mathbf{N}}_{x,\text{C}} & \mathbf{0} \\
       \mathbf{D}_{x_{d,\text{C}}}\bar{\mathbf{N}}_{x\theta,\text{C}} & \bar{\mathbf{N}}_{y,\text{C}}
    \end{bmatrix}^\dagger
    \Big(
    \begin{bmatrix}
       \mathbf{x}_{d,\text{C}} \\
       \mathbf{y}_{d,\text{C}}
    \end{bmatrix}
    - \\
    \begin{bmatrix}
       \bar{\mathbf{N}}_{x,\text{PC}} & \mathbf{0} \\
       \mathbf{D}_{x_{d,\text{PC}}}\bar{\mathbf{N}}_{x\theta,\text{PC}} & \bar{\mathbf{N}}_{y,\text{PC}}
    \end{bmatrix}
    \begin{bmatrix}
       \bar{\mathbf{p}}_{x,\text{P}} \\
       \bar{\mathbf{p}}_{y,\text{P}} 
    \end{bmatrix}
    \Big)
\end{multline}
The coefficients in the decoupled approximation can be calculated sequentially. First, we calculate $\bar{\mathbf{p}}_{x,\text{C}}$ using $\bar{\mathbf{p}}_{\text{C}}$ in Eq. (39) (29) applied to the $x$-axis, and use the obtained coefficients to obtain $\bar{\mathbf{p}}_{y,\text{C}}$ as
\begin{multline}
    \bar{\mathbf{p}}_{y,\text{C}} = \bar{\mathbf{N}}_{y,\text{C}}^\dagger\Big(\mathbf{y}_{d,\text{C}} - \\ (\mathbf{D}_{x_{d,\text{PC}}}\bar{\mathbf{N}}_{x\theta,\text{PC}}\bar{\mathbf{p}}_{x,\text{P}} + \mathbf{D}_{x_{d,\text{C}}}\bar{\mathbf{N}}_{x\theta,\text{C}}\bar{\mathbf{p}}_{x,\text{C}}) - \bar{\mathbf{N}}_{y,\text{PC}}\bar{\mathbf{p}}_{y,\text{P}}\Big).
\end{multline}

\section*{ACKNOWLEDGMENT}

This work was supported in part by the National Science Foundation [grants numbers DGE 1256260 and CMMI 1825133]. The authors would like to thank Mr. Chandler Harris for his assistance in retrofitting the H-frame 3D printer used in the experiments.


\end{document}